\documentclass[twocolumn,amsmath,amssymb,floatfix,prl,showpacs,footinbib,natbib]{revtex4}
\usepackage{amsmath,amssymb,bm,graphicx,url,psfrag}

\usepackage{color}

\newcommand{\be}{\begin{equation}}
\newcommand{\ee}{\end{equation}}

\DeclareMathOperator{\rre}{Re}

\begin{document}
\title{Controlling the spontaneous emission of a superconducting transmon qubit}
\author{A.\ A.\ Houck}
\affiliation{Departments of Physics and Applied Physics, Yale University, New Haven, Connecticut 06520, USA}
\author{J.\ A.\ Schreier}
\affiliation{Departments of Physics and Applied Physics, Yale University, New Haven, Connecticut 06520, USA}
\author{B.\ R.\ Johnson}
\affiliation{Departments of Physics and Applied Physics, Yale University, New Haven, Connecticut 06520, USA}
\author{J.\ M.\ Chow}
\affiliation{Departments of Physics and Applied Physics, Yale University, New Haven, Connecticut 06520, USA}
\author{Jens Koch}
\affiliation{Departments of Physics and Applied Physics, Yale University, New Haven, Connecticut 06520, USA}
\author{J.\ M.\ Gambetta}
\affiliation{Departments of Physics and Applied Physics, Yale University, New Haven, Connecticut 06520, USA}
\author{D.\ I.\ Schuster}
\affiliation{Departments of Physics and Applied Physics, Yale University, New Haven, Connecticut 06520, USA}
\author{L.\ Frunzio}
\affiliation{Departments of Physics and Applied Physics, Yale University, New Haven, Connecticut 06520, USA}
\author{M.\ H.\ Devoret}
\affiliation{Departments of Physics and Applied Physics, Yale University, New Haven, Connecticut 06520, USA}
\author{S.\ M.\ Girvin}
\affiliation{Departments of Physics and Applied Physics, Yale University, New Haven, Connecticut 06520, USA}
\author{R.\ J.\ Schoelkopf}
\affiliation{Departments of Physics and Applied Physics, Yale University, New Haven, Connecticut 06520, USA}

\begin{abstract}
We present a detailed characterization of coherence in seven transmon qubits in a circuit QED architecture.  
We find that spontaneous emission rates are strongly influenced by far off-resonant modes of the cavity and can be understood within a semiclassical circuit model.
A careful analysis of the spontaneous qubit decay into a microwave transmission-line cavity can accurately predict the qubit lifetimes over two orders of magnitude in time and more than an octave in frequency.   Coherence times $T_1$ and $T_2^*$ of more than a $\mu\rm{s}$ are reproducibly demonstrated.
\end{abstract}
\pacs{03.67.Lx, 85.25.-j, 42.50.-p}
\date{March 31, 2008}
\maketitle

Coherence poses the most important challenge for the development of a superconducting quantum computer. As the dephasing time $T_2^*$ can never exceed twice the relaxation time $T_1$, it is the relaxation time which ultimately  sets the limit on qubit coherence. Although $T_2^*$ turned out to be small compared to $T_1$ in the earliest superconducting qubits \cite{nakamura}, steady progress over the last decade has significantly reduced this gap \cite{vion2,wallraff,bertet3,steffen,schreier}. Recently, the transmon, a new type of qubit immune to $1/f$ charge noise, has been shown to be nearly homogeneously broadened ($T_2^*\simeq2T_1$) \cite{schreier}.    Therefore, understanding relaxation mechanisms is becoming critical to further improvements in both $T_1$ and $T_2^*$.  Progress in this direction will be based on the  accurate modeling of contributions to $T_1$ and the reliable fabrication of many qubits reaching consistent coherence limits.  

One of the main advantages of superconducting qubits is their strong interaction with the wires of an electrical circuit, making their integration with fast control and readout possible and allowing for large, controllable couplings between widely separated qubits \cite{schoelkopfnature}.  The large coupling also implies a strong interaction between the qubits and their electromagnetic environment, which can lead to a short $T_1$. However, careful control of the coupling to the environment has been shown to allow prevention of circuit dissipation \cite{esteve,devoret3}.
Relaxation times have been studied in a wide variety of superconducting qubits, created with different fabrication techniques, and measured  with a multitude of readout schemes. Typically,  values of $T_1$ vary strongly from sample to sample as they can depend on many factors including materials, fabrication, and the design of both readout and control circuitry. In some instances a separation of these components has been achieved \cite{bertet2,martinis2,neeley,astafiev}, but typically it is difficult to understand the limiting factors, and $T_1$ often varies strongly even among nominally identical qubit samples.

Here, we demonstrate that in a circuit quantum electrodynamics (QED) architecture, where qubits are embedded in a  microwave transmission line cavity\cite{blais1,wallraff}, transmon qubits have reproducible and understandable relaxation times. Due to the simple and well-controlled fabrication of the qubit and the surrounding circuitry, involving only two lithography layers and  a single cavity for both control and readout, we are able to reliably understand and predict qubit lifetimes. This understanding extends to a wide variety of different qubit and cavity parameters. We find excellent agreement between theory and experiment for seven qubits over two orders of magnitude in relaxation time and more than an octave in frequency. The relaxation times are set by either spontaneous emission through the cavity, called the Purcell effect \cite{purcell}, or a shared intrinsic limit  consistent with a lossy dielectric.  
 Surprisingly, relaxation times are often limited by electromagnetic modes of the circuit which are far detuned from the qubit frequency. In the circuit QED implementation studied here, the infinite set of cavity harmonics reduces the Purcell protection of the qubit at frequencies above the cavity frequency.

Generally, any discrete-level system coupled to the continuum of modes of the electromagnetic field is subject to radiative decay.  By placing an atom in a cavity, the rate of emission can be strongly enhanced \cite{purcell,goy} or suppressed \cite{kleppner,hulet, jhe}, depending on whether the cavity modes are resonant or off-resonant with the emitter's transition frequency. This effect is named after E.\ M.\ Purcell \cite{purcell}, who considered the effect of a resonant electrical circuit on the lifetime of nuclear spins.   Suppression of spontaneous emission provides effective protection from radiative qubit decay in the dispersive regime, where qubit and cavity are detuned \cite{blais1}. Specifically, the Purcell rate for dispersive decay is given by $\gamma_\kappa=(g/\Delta)^2\kappa$, where $g$ denotes the coupling between qubit and cavity mode, $\Delta$ their mutual detuning, and $\kappa$ the average photon loss rate.

\begin{figure}
    \centering
        \includegraphics[width=0.95\columnwidth]{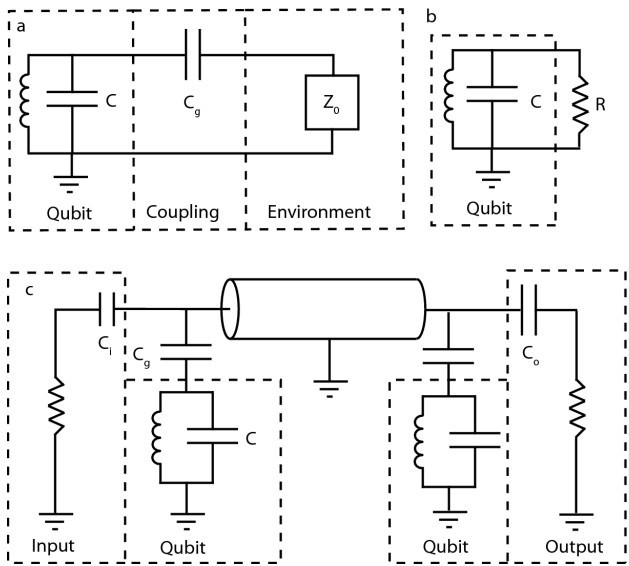}
    \caption{Circuit model of qubit relaxation. (a) Generalized model for a qubit coupled to an environment. (b) Reduced model of dissipation. The coupling capacitor and environment impedance are replaced by an effective resistance $R=1/\rre[Y(\omega)]$,  where $Y(\omega)$ is the admittance of the rest of the circuit seen by the qubit. The $T_1$ for the qubit is $RC$, where $C$ is the qubit capacitance.  (c) Full circuit diagram.  Qubits are capacitively coupled to either end of a transmission line cavity.  Both the input and output of the cavity are connected to a $50\,\Omega$ environment.  The cavity is asymmetric in the sense that the input capacitance is smaller than the output capacitance. 
\label{fig:fig1}}
\end{figure}

The suppression and enhancement of decay rates can alternatively be calculated within a circuit model. For concreteness, we consider the case of a qubit capacitively coupled to an arbitrary environment with impedance $Z_0(\omega)$, see Fig.\ \ref{fig:fig1}(a). This circuit may be reduced to a qubit coupled to an effective dissipative element, see Fig. \ref{fig:fig1}(b). Specifically, replacing the coupling capacitor $C_g$ and the environment impedance $Z_0$ by an effective resistor $R=1/\rre[Y(\omega)]$, one finds\cite{esteve,devoret3} that the $T_1$ is given  by $RC$, where $C$ is the qubit capacitance.  Choosing a purely resistive environment, $Z_0=50\,\Omega$, yields a decay rate $\gamma\simeq\omega^2 Z_0 C_g^2/C$.  If instead we couple to a parallel $LRC$ resonator, the calculated radiation rate can be reduced to that of the atomic case, $\gamma_\kappa=(g/\Delta)^2\kappa$, thus reproducing the Purcell effect.

\begin{figure}
    \centering
        \includegraphics[width=0.97\columnwidth]{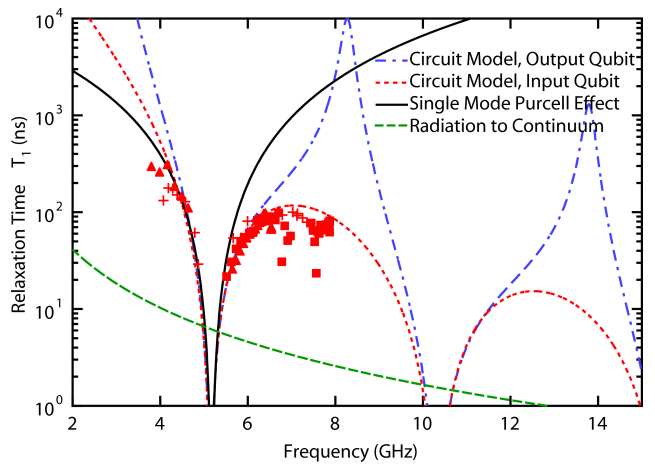}
    \caption{Comparison of circuit and single-mode models of relaxation.  Spontaneous emission lifetimes into a single-mode cavity are symmetric about the cavity frequency, while within the circuit model lifetimes below the cavity are substantially longer than above.  The measured $T_1$ for three similar qubits deviates substantially from the single-mode prediction, but agrees well with the circuit model.  The expected decay time for radiation into a continuum is shown for comparison.
\label{fig:fig2}}
\end{figure}

The qualitative features of the Purcell effect are apparent in measurements of $T_1$,  shown  for 3 qubits in Fig.\ \ref{fig:fig2}, measured with a dispersive readout by varying a delay time between qubit excitation and measurement \cite{wallraff2,schreier}. Near the cavity resonance at $5.2\,\text{GHz}$, spontaneous emission is Purcell-enhanced and $T_1$ is short. Away from resonance, the cavity protects the qubit from decay and the relaxation time is substantially longer than expected for decay into a continuum. However, at detunings above the cavity frequency, the measured $T_1$  deviates significantly from the single-mode Purcell prediction. This deviation can be directly attributed to the breakdown of the single-mode approximation. 

The cavity does not just support a single electromagnetic mode, but also all higher harmonics of the fundamental mode. This has a striking impact on relaxation times.  At first glance, it would appear that the effects of higher modes could be ignored when the qubit is close to the fundamental frequency and detuned from all higher modes.  However, the coupling $g_n$ to the $n^{\rm{th}}$ mode of the cavity increases with mode number, $g_n = g_0 \sqrt{n+1}$.  In addition, the input and output capacitors act as frequency-dependent mirrors, so that the decay rate of the $n^{\rm{th}}$ harmonic, $\kappa_n=(n+1)^2\kappa$, is larger than that of the fundamental.  As a result, higher modes significantly contribute to the qubit decay rate, and the simple single-mode quantum model turns out to be inadequate for understanding the $T_1$ of the system.  
The naive attempt to treat the fundamental and harmonics in terms of a multi-mode Jaynes-Cummings Hamiltonian faces problems with divergences. Work on developing a consistent quantum model is currently under way \cite{gambettawork}.

Here, we follow the alternative route of calculating $T_1$ semiclassically, on the basis of the full underlying circuit, and show that this accurately reproduces the measured $T_1$. The relationship between the classical admittance $Y(\omega)$ of a circuit  and its dissipation has long been known \cite{esteve,devoret3}, providing a practical means of understanding relaxation rates \cite{neeley}.  The full calculation includes a transmission line cavity rather than a simple $LRC$ resonator, see Fig.\ \ref{fig:fig1}(c). The results from this  are shown in Fig.\ \ref{fig:fig2}, and
 reveal two striking differences as compared to the single-mode model: First, there is a strong asymmetry between relaxation times for qubit frequencies above (positive detuning) and below (negative detuning) the fundamental cavity frequency. While the single-mode model predicts identical relaxation times for corresponding positive and negative detunings, $T_1$ can be two orders of magnitude shorter for positive detunings than for negative detunings in the circuit model.  Second,
 the circuit model shows a surprising dependence of $T_1$ on the qubit position in the cavity. 
 While qubits located at opposite ends of the cavity have the same $T_1$ within the single-mode model, the circuit model correctly captures the asymmetry induced by the differing input and output coupling capacitors and leads to vastly different $T_1$. The circuit model accurately resolves the discrepancy between the experimental data and the single-mode model, see Fig.\ \ref{fig:fig2}.

\begin{table}
\begin{tabular}{lccccc}
  ID & Res. & $\omega_r\,\rm{(GHz)}$ & $\kappa\,\rm{(MHz)}$ & $g\,\rm{(MHz)}$ & Pos.\\
  \hline
  1  & Al on Si & $5.17$ & $44$ & 107& In \\
  2L & Al on Si & $5.19$ & $33$ & 105& In \\
  2R & Al on Si & $5.19$ & $33$ & 105& Out \\
  3L & Nb on Sapph & $6.69$ & $40$ & 166& In \\
  3R & Nb on Sapph & $6.69$ & $40$ & 50& Out \\
  4L & Nb on Sapph & $6.905$ & $0.7$ & 150& In \\
  4R & Nb on Sapph & $6.905$ & $0.7$ & 55& Out \\
  \hline
\end{tabular}
\caption{Qubit parameters.  Sample 1 is a single-qubit sample, all others are two-qubit samples.  The Res.\ column indicates material and substrate for the cavity.  The Pos.\ column indicates the position of the qubit at the input or output end of the cavity.}
\end{table}

The predictive power of the circuit model extends to all of our transmon qubits. Here, we present $T_1$ measurements on a representative selection of seven qubits. The qubits were fabricated on both oxidized high-resistivity silicon and sapphire substrates, and coupled to microwave cavities with various decay rates and resonant frequencies.  Table 1 provides parameters for each of the seven qubits.  Qubits are fabricated via electron beam lithography and a double-angle evaporation process ($25\,\rm{nm}$ and $80\,\rm{nm}$ layers of aluminum), while cavities are fabricated by optical lithography with either lifted-off Al or dry-etched Nb on a Si or sapphire substrate \cite{frunzio}.

\begin{figure}
    \centering
    \includegraphics[width=0.97\columnwidth]{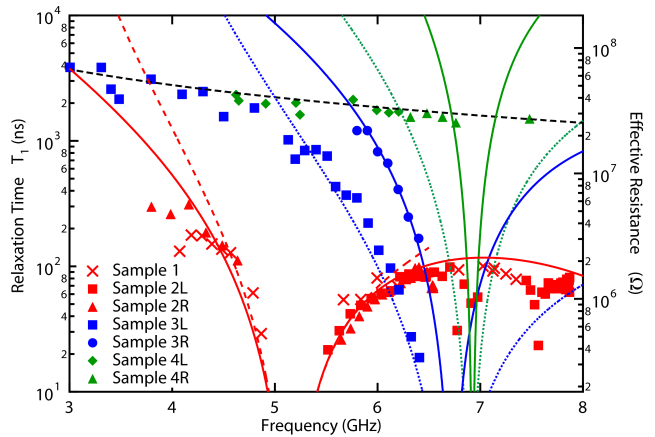}
    \caption{Relaxation times for seven superconducting qubits.  Predictions for qubit lifetime based on the circuit model (colored lines) agree well with observed relaxation times (points).  Solid lines represent predictions for input side (L) qubits, while dashed lines correspond to output side (R) qubits.  All sapphire qubits (blue and green) reach the same common intrinsic limits (black line), with lifetimes limited to a constant $Q\sim70,000$.  Some deviation is seen in the lowest frequency silicon qubits, though it is unclear if this is an intrinsic limit.  Qubit lifetimes are accurately predicted over a wide range of frequencies and more than two orders of magnitude in time.  
\label{fig:fig3}}
\end{figure}

Predictions from the circuit model are in excellent agreement with observed qubit lifetimes (see Fig.\ \ref{fig:fig3}), up to a $Q = 70,000$ for qubits on sapphire.  The agreement is valid over more than two orders of magnitude in qubit lifetime and more than an octave of frequency variation.  We emphasize that the circuit model does not correspond to a fit to the data, but rather constitutes a prediction based on the independently measured cavity parameters $\omega_r$ and $\kappa$, and the coupling $g$. 

In the qubits on silicon, coherence times of no more than $100\,\rm{ns}$ are observed above the cavity resonance, far below predictions from the single-mode model, but consistent with the circuit model.  Initially, this caused concern for the transmon qubit:  it appeared as if the transmon solved the $1/f$-noise dephasing problem for charge qubits, but introduced a new relaxation problem \cite{schuster,majer,houck1}.  However, with the circuit model of relaxation, it is now clear that the $100\,\rm{ns}$ limit originated from the surprisingly large spontaneous emission rate due to higher cavity modes.  By working at negative detunings instead, it is possible to achieve long  relaxation times, here observed up to $4 \mu\rm{s}$.  

All qubits on sapphire substrates reach a shared intrinsic limit of $Q = 70,000$ when not otherwise Purcell limited. The constant-$Q$ frequency dependence of the intrinsic limit ($T_1 \propto 1/\omega$) is suggestive of dielectric loss as the likely culprit.    While the observed loss tangent $\tan{\delta} \sim 10^{-5}$ is worse than can be achieved for sapphire, it is not unreasonable depending on the type and density of surface dopants that might be present\cite{braginsky1,braginsky2}.  The overall reproducibility of the intrinsic limit gives hope that future experiments may isolate its cause and reveal a solution.  It is instructive to reexpress the relaxation times in terms of a parasitic resistance, see Fig.\ \ref{fig:fig3}. Note that here a $T_1$ of a microsecond roughly corresponds to a resistance of $20\,\rm{M}\Omega$.  To build more complex circuits with still longer $T_1$, all dissipation due to parasitic couplings must be at the $\rm{G}\Omega$ level.


\begin{figure}	
    \centering
    \includegraphics[width=0.97\columnwidth]{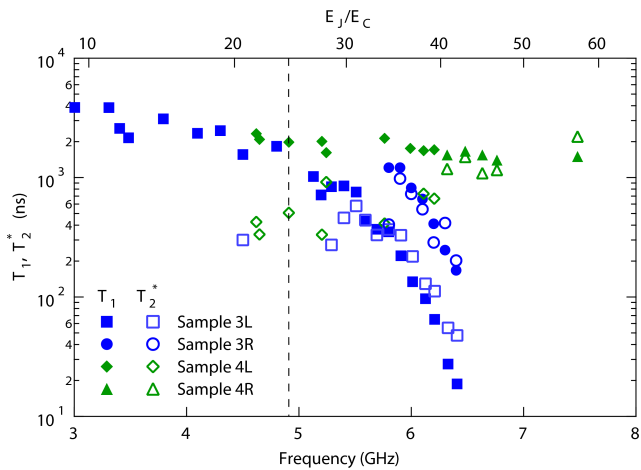}
    \caption{Dephasing times for four sapphire qubits.  Measured dephasing times for each of the four sapphire qubits are nearly homogenously broadened, with $T_2^*$ (open symbols) similar to $T_1$ (closed symbols) over a wide range of frequencies, even away from the flux sweet spot (the maximum frequency for each qubit). Charge noise, is suppressed exponentially in the ratio of Josephson to charging energies $E_J/E_C$ (top axis), tuned along with qubit frequency (bottom axis) by changing an applied magnetic field.  For small $E_J/E_C$  charge noise dephasing is relevant and causes short $T_2^*$.  Onset of significant charge noise is indicated by the dashed vertical line.
\label{fig:fig4}}
\end{figure}

Transmon qubits benefit greatly from the increased relaxation times, as they are insensitive to $1/f$-charge noise, the primary source of dephasing in other charge qubits. As a result, coherence is limited primarily by energy relaxation  and transmons are nearly homogeneously broadened ($T_2^* \simeq 2T_1$). Improvements in $T_1$ thus translate directly into improvements in dephasing times $T_2^*$. This is demonstrated in Fig.\ \ref{fig:fig4}, showing a comparison of relaxation and dephasing times. Here, $T_2^*$ is measured in a pulsed Ramsey experiment and without echo \cite{schreier}.  The gain in coherence time is most striking in samples with a higher-frequency cavity, $\omega_r/(2\pi)\sim 7\,\rm{GHz}$, where it is easier to operate at negative detunings and attain long $T_1$. In all these samples, we observe consistently long dephasing times of nearly a microsecond, with the largest $T_2^*$ exceeding two microseconds without echo.

There are two main effects determining the observed dependence of $T_2^*$ on the qubit frequency. First, away from the maximum frequency for each qubit, i.e.\ the flux sweet spot \cite{vion2}, the sensitivity to flux noise increases. This can cause additional inhomogeneous broadening. Despite this, $T_2^*$ remains close to two microseconds, even away from the flux sweet spot. Second, tuning the qubit frequency via $E_J$ directly affects the ratio of Josephson to charging energy, $E_J/E_C$, which dictates the sensitivity to charge noise.  At low qubit frequencies, the qubits regain the charge sensitivity of the Cooper Pair Box, thus explaining the strong drop in dephasing times seen in Fig.\ \ref{fig:fig4}.

Future improvements in $T_2^*$ require further improvements in $T_1$.  The accurate modeling of relaxation processes will be essential as quantum circuits become more complicated. In particular, the addition of multiple cavities and individual control lines may introduce accidental electromagnetic resonances. As we have shown here, even far off-resonant modes of a circuit can have a dramatic impact on qubit lifetimes. However, with careful circuit design, it should be possible not only to avoid additional accidental resonances, but to utilize the circuit model of relaxation to build filters to minimize dissipation.
The concise understanding of spontaneous emission lifetimes in our system, and the reproducibility of intrinsic lifetimes open up vistas for a  systematic exploration of limits on coherence.

\begin{acknowledgments}
This work was supported in part by Yale University via a Quantum Information and Mesoscopic Physics Fellowship (AAH, JK), by CNR-Istituto di Cibernetica (LF), by LPS/NSA under ARO Contract No.\ W911NF-05-1-0365, and
the NSF under Grants Nos.\ DMR-0653377 and DMR-0603369. 
\end{acknowledgments}


\end{document}